\documentclass{bmcart}
\usepackage{graphicx}
\usepackage[utf8]{inputenc} %unicode support
\usepackage{amssymb,amsfonts,amsmath,verbatim}

\startlocaldefs

\endlocaldefs

\begin{document}
\begin{frontmatter}

\begin{fmbox}
\dochead{Sentiment cascades in the 15M movement}

\title{Sentiment cascades in the 15M movement}

\author[
   addressref={aff1},
]{\inits{RA}\fnm{Raquel} \snm{Alvarez}}

\author[
   addressref={aff2},                   
   corref={aff2},                       
   email={dgarcia@ethz.ch}  
]{\inits{DG}\fnm{David} \snm{Garcia}}

\author[
   addressref={aff1},
]{\inits{YM}\fnm{Yamir} \snm{Moreno}}

\author[
   addressref={aff2},
]{\inits{FS}\fnm{Frank} \snm{Schweitzer}}

\address[id=aff1]{
  \orgname{Institute for Biocomputation and Physics of Complex Systems, University of Zaragoza}, 
  \street{Campus Rio Ebro},                     
  \postcode{50018}                                
  \city{Zaragoza},                             
  \cny{Spain}                                   
}
\address[id=aff2]{%
  \orgname{Chair of Systems Design, ETH Zurich},
  \street{Weinbergstrasse 56/58},
  \postcode{8092}
  \city{Zurich},
  \cny{Switzerland}
}

\begin{artnotes}
%\note{Sample of title note}     % note to the article
%\note[id=n1]{Equal contributor} % note, connected to author
\end{artnotes}

\end{fmbox}

\begin{abstractbox}
\begin{abstract} 
Recent grassroots movements have suggested that online social networks might
play a key role in their organization, as adherents have a fast, many-to-many,
communication channel to help coordinate their mobilization. The structure and
dynamics of the networks constructed from the digital traces of protesters
have been analyzed to some extent recently. However, less effort has been
devoted to the analysis of the semantic content of messages exchanged during
the protest.  Using the data obtained from a microblogging service during the
brewing and active phases of the 15M movement in Spain, we perform the first
large scale test of theories on collective emotions and social interaction in
collective actions. Our findings show that activity and information cascades
in the movement are larger in the presence of negative collective emotions and
when users express themselves in terms related to social content. At the level
of individual participants, our results show that their social integration in
the movement, as measured through social network metrics, increases with their
level of engagement and of expression of negativity. Our findings show that
non-rational  factors play a role in the formation and activity of social
movements through online media, having important consequences for viral
spreading. 
\end{abstract}


\end{abstractbox}

\end{frontmatter}

\section{Introduction}

The Occupy and 15M movements are recent examples of self-organized social
movements that appeared in  developed countries in response to a widespread
perception of social and economical inequality \cite{Hughes2011,
Castaneda2012}. While these movements address a wide range of problems in
different countries, they share a common factor, their usage of social media
to communicate, organize, and deliberate about the purpose of the movement and
its actions \cite{Zuckerman2014, Tufekci2014}.  Social media allow the
participants of these movements to circumvent their lack of influence on
state-and private-owned mass media \cite{Herman2008}, creating an emergent
structure without a central actor or decision group. As a side effect,  these
movements leave public digital traces of their activity, which allows us to
analyze their formation, behavior, and organization up to unprecedented scales
and resolutions.

Collective actions pose a classical paradox of the tragedy of the commons
\cite{Olson2009}: A purely rational individual would choose not to participate
in a movement it agrees with, as it would receive its collective benefits
without the associated costs and risks of taking part on it.  Thus, the
existence of collective actions and social movements requires considerations
beyond rational decisions, including emotions \cite{Garcia2013} and social
influence \cite{Mavrodiev2013} between the participants of a social movement.
In this article, we present a detailed quantitative analysis of the digital
traces of the 15M movement, the Spanish precursor of Occupy movements across
Europe and America \cite{Castaneda2012}. This decentralized movement emerged in Spain in the aftermath of the so-called Arab Spring as a reaction to public spending cuts and the economic crisis.  It was mainly nucleated in online social networks before massive offline demonstrations ended up in several camp sites in many city squares. From that point on, the movement consolidated and lasted for months. Even today the foundations of the 15M movement drives the political agenda of some new parties and associations in Spain. 

Our analysis covers its online social structure and the content of the public messages exchanged in
\texttt{Twitter}, the main online medium used by the movement.
\texttt{Twitter} users create directed links to follow the messages of other
users and communicate through short public messages called \emph{tweets}. We
analyze the content of a large set of tweets about the 15M movement,
extracting sentiment values and semantic content related to  social and
cognitive processes. Our aim is to explore the role of social emotions in
group activity and collective action. We address how emotional interaction
supports the creation of social movements and how emotional expressions lead
to the involvement of the participants of the movement.

According to the theory of collective identity of Emile Durkheim, group
gatherings contribute to the creation of collective identity by means of
rituals and symbols that produce an atmosphere of emotional synchrony
\cite{Durkheim1915}. These rituals are often emotionally charged and show  an
inverse relation between emotional intensity and frequency
\cite{Atkinson2011}.  The emotions experienced by the participants of these
gatherings contribute to social inclusion and identification with the
collective, as empirically shown in a variety of experiments \cite{Paez2013}.
This also holds for the 15M movement, for which survey results show that
participants of the large demonstrations across Spain in 2011 felt a stronger
emotional communion with the movement, in comparison to those participants who
did not attend to  the demonstration \cite{Paez2013b}. In this article, we
provide a quantitative analysis of how collective identity and action emerged
in the 15M movement, through the analysis of the digital traces of  its
participants in the \texttt{Twitter} social network. We pay special attention
to emotional expression in tweets, social inclusion in the follower network of
the participants of the movement, and sentiment polarization in the creation
and social response to the movement.

Online media offer large datasets to explore political activity at a large
scale, to find out about popularity and mobilization in political campaigns
\cite{Garcia2012}, and political alignment based on public messages
\cite{Conover2011, Garcia2013c}. Analyzing online social networks, for example
by means of the k-core decomposition method, can also reveal  relevant
information about  the role of influential individuals
\cite{Kitsak2010,Banos2013}  and the social resilience of an online community
\cite{Garcia2013b}.   Users of online social networks communicate through
public messages that provide the breeding ground for collective emotional
states, which have the potential to create the identity and mobilization of
the movement. Previous states of collective emotional persistence were
detected in the short messages of IRC chats \cite{Garas2012} and spread
through social networks as cascades of emotions \cite{Suvakov2013}, forming
patterns in which happy individuals are likely to be connected to other happy
ones \cite{Bollen2011}.

In our analysis we follow a top-down approach, from the collective level of
the movement to the actions of its individuals and their relations. We start
by analyzing the dynamic aspect of the 15M movement, identifying cascades of
tweets as in previous research \cite{Banos2013}.  We measure the size of these
cascades in terms of the amount of participants communicating in the cascade
(spreaders), and the amount of participants exposed to the cascade
(listeners).   We analyze how cascade sizes depend on  collective emotions and
the use of terms related to cognitive and social processes. Finally, we zoom
into the microscopic level of individuals and their interactions, creating an
additional dataset of tweets of each participant of the 15M movement. We
relate their expression of emotions, cognitive, and social processes to their
activity and social integration in the movement, as quantified by their k-core
centrality within the social network.

\section{Results}

\subsection{Sentiment analysis in Spanish}

Our adaptation of SentiStrength to the Spanish language \cite{Thelwall2013},
explained in the Materials and Methods section, reaches accuracy values above
$0.6$ for two test datasets (See SI table II). These results are comparable to
state of the art unsupervised techniques of sentiment analysis for the Spanish
language \cite{Diaz2013}. Furthermore, the quality of the sentiment analysis
tool does not differ for tweets related to politics and economics (See SI
table III). This  result shows that our application of SentiStrength is valid,
as 15M tweets appear in a context of political protests related to economic
measures.

\subsection{Activity and information cascades}

Previous research has shown a positive relation between retweeting and
emotional content \cite{Pfitzner2012}.  Here, we go beyond  a plain retweeting
behavior and analyze cascades  associated with the 15M topic. 
We quantify emotions in the tweets related to the 15M movement through
sentiment analysis  on a dataset of tweets selected by the hashtags related to
the 15M movement, as explained in the Materials and Methods section. 
We focus on the analysis of  tweet cascades, also defined in
Materials and Methods,  to detect how the content of tweets influences both
the activity and the volume  of information perceived by the participants of
the movement.  

We define the size of an activity cascade as the number of
unique \texttt{Twitter} users that produce a tweet in the cascade, also known
as the number of spreaders, $n_{sp}$. The associated size of an information
cascade corresponds to the amount of unique users who receive some  tweet of
the cascade in their tweet feeds. This concept, commonly known as exposure of
the tweets in the cascade, is the sum of the amount of participants who follow
at least one spreader, denoted as $n_c$.

We characterize the collective emotions in a cascade, $c$, using the ratios
of positive, neutral, and negative tweets:
\begin{equation}
r_p(c) = \sum_{m \in T(c)}\frac{\Theta(e_m==1)}{N(c)} \qquad r_u(c) =
\sum_{m \in T(c)}\frac{\Theta(e_m==0)}{N(c)} \qquad
r_n(c) = \sum_{m \in T(c)}\frac{\Theta(e_m==-1)}{N(c)}
\label{eq:emo}
\end{equation}
where $T(c)$ is the set of tweets on cascade $c$, $e_m$ is the emotional
content of tweet $m$ as given by the sentiment analysis tool, and $N(c)$ is
the total amount of tweets related to the 15M movement comprising cascade $c$.
The collective emotions expressed by the participants of a cascade have the
potential to activate additional participants, influencing its activity
size and information spreading. To test this possibility, we classify cascades
according to their ratios of positive, $r_p(c)$, and negative, $r_n(c)$,
tweets and  compare these  with the total ratios of positive tweets, $\mu_p$,
and negative tweets, $\mu_n$. If both $r_p(c) \leq \mu_p$ and $r_n(c) \leq
\mu_n$, we label the cascade as \emph{neutral}. If $r_p(c) > \mu_p$ and
$r_n(c) \leq \mu_n$, we label it as \emph{positive},  and if $r_n(c) > \mu_n$
and $r_p(c) \leq \mu_p$, we label it as \emph{negative}. When both  $r_p(c) >
\mu_p$ and $r_n(c) > \mu_n$, we label it as \emph{bipolar}. From the total of
$96065$ cascades we analyzed, $43415$ are positive ($45.19 \%$), $20989$ are
negative ($21.85 \%$), $30664$ are neutral ($31.92\%$), and $997$ are bipolar
($1.04 \%$).

  \begin{figure}[ht]
  \includegraphics[angle=-90,width=0.95\textwidth]{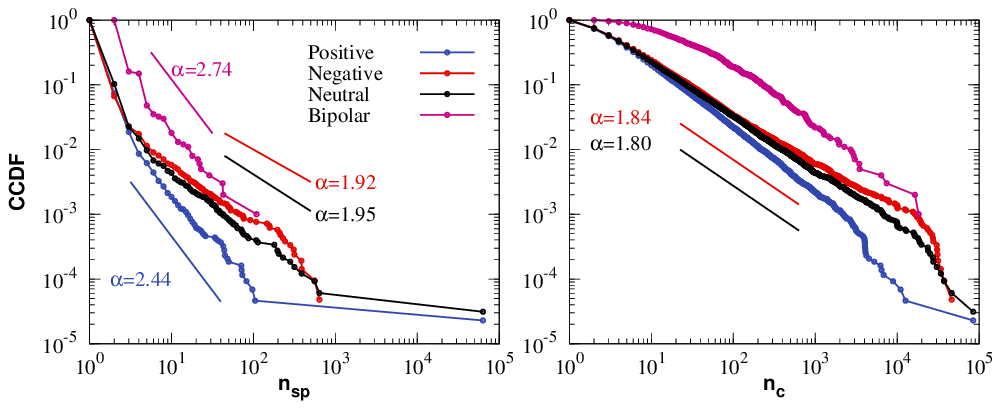}
  \caption{Complementary cumulative density function for activity
cascade sizes (left) and information cascade sizes (right). In this case,
cascades have been classified according to their aggregate sentiment into
positive, negative, neutral and bipolar.\label{fig:casc_agregated_sentiment}}
 \end{figure}

Figure~\ref{fig:casc_agregated_sentiment} shows the complementary cumulative
density function (CCDF) of activity cascade sizes, $P(x>n_{sp})$, and of
information cascade sizes, $P(x>n_{c})$. Bipolar cascades are likely to be
larger than positive, negative, and neutral, but  we do not observe extremely
large bipolar cascades, since they are less frequent in general.  We apply
the  Kolmogorov-Smirnov test  (KS)  with a tail correction factor, as
explained in \cite{Clauset2009}, to test the equality of information and
activity cascades across emotion classes. The KS tests validate the
observation that cascade sizes (both $n_{sp}$ and $n_c$) in bipolar cascades
are different from in any of the other three classes (details in SI Table V).
Furthermore, the test rejects the null hypothesis that positive cascade sizes
are distributed as their negative and neutral counterparts, and only fails to
reject the null hypothesis for the case of negative versus neutral information
cascades.

To further compare these cascades classes, we fit power law distributions
of the form $p(x) \sim x^{-\alpha}$  for $x \geq x_{min}$, to the empirical
distributions of $n_{sp}$ and $n_c$.   The power law distribution is
characterized by a skewed right tail that starts at a minimum value of
$x_{min}$ and scales with exponent $\alpha$.  The estimated value of $\alpha$
can reveal important properties of how the mean and variance of the
distribution scale with system size, which in our case is the amount of users
in the network. For example, $\alpha \leq 2$ implies that both the mean and
the variance of $x$ increase with the size of the sample \cite{Newman2005},
and thus the expected cascade size would increase for larger movements. Power 
law distributions were fitted using the Python package \textit{powerlaw}
\cite{Alstott2014}.  Power law fits reveal  that $n_{sp}$ for positive
cascades decays with an exponent $\alpha=2.44\pm 0.07$ (see Table VI of the SI
for details). This means that the distribution of $n_{sp}$ decreases faster
than negative and neutral cascades,  with exponents $\alpha=1.92\pm 0.07$ and
$\alpha=1.95\pm 0.09$ respectively, but slower than bipolar ones, which are
best fitted with an exponent of $\alpha=2.74\pm 0.32$. The exponents of
positive and bipolar activity cascade sizes, right above $2$, imply that their
expected size  does not scale with system size, i.e. they do not become larger
with  larger populations. This is not the case for negative and neutral
activity cascades, with exponents too close to 2  to arrive at any conclusion.

We also  investigate the goodness of the fits by comparing them to fits to
other distributions.  In this way we are able to identify if a power law
behavior is a good description of  our data. Specifically, we calculate the
likelihood ratio, $R$ (see SI Table VI), between the  power law and a
lognormal  distribution, and the corresponding $p$-value indicating the
significance for the observed likelihood direction. Positive  values of $R$
suggest that the most likely model is a  power law distribution. However, when
these values are obtained in combination with high $p$-values ($p>0.05$),  the
evidence of a power law versus a lognormal distribution is moderated
\cite{Alstott2014}.

For the case of information cascades, the distributions of negative and
neutral cascade sizes are similar  (KS p-value $0.285$),  but the null
hypothesis that they have the same size as the positive ones could be
rejected. The fit of  power-law distributions reveals that the scaling of
positive and bipolar cascade sizes are similar, $\alpha=2.01\pm 0.02$ and
$\alpha=1.99\pm 0.08$ respectively, while negative and neutral information
cascade size distributions decay with $\alpha=1.80\pm 0.01$ and
$\alpha=1.84\pm 0.01$ (see SI Table VI for details). The exponents below 2
imply that the expected size of the audience of negative and neutral cascades
increases with system size, while bipolar and positive have exponents too
close to $2$ to arrive to any conclusion. Furthermore, the log-likelihood
analysis indicates that data is better described by a log-normal distribution
for positive and bipolar cascades.  Although this evidence is  moderated
($p$-values $0.34$ and $0.67$ respectively), it suggests that positive and
bipolar information cascades  are more likely to follow a lognormal
distribution than a power law distribution and thus do not scale with system
size.

The above results highlight the role of emotional expression inside a social
movement: cascades with positive emotions (including bipolar ones) do not seem
to trigger more activity nor spread  more information than those with more
objective and negative expression. This difference for distinct collective
emotions  opens the question of the role of the first tweet in the cascade. To
test if this effect is due to the sentiment of the first tweet in the
cascade, we extend our analysis to compare the distributions of cascades
sizes for cascades that started with a positive, negative, or a neutral tweet
(SI section IV). We find no consistent differences  on cascade sizes
depending on the emotions expressed in the first tweet of the cascade. This
highlights the role of collective emotions in spreading processes: it is not
the emotion of the tweet that triggers the cascade what matters, it is the
overall sentiment of all the people involved in the cascade.

In addition to the sentiment, the semantic content of tweets can be analyzed
with respect to social and cognitive content through psycholinguistic methods
(see Materials and Methods). In particular, the content of tweets in relation
to social and cognitive processes have the potential to determine the success
or failure of the spreading process. To check this we perform an analogous
analysis of the distribution of the cascade size for different cascades types.
In particular, we  apply the same method as in the previous section to
classify the collective emotions in cascades, comparing their ratios of
social and cognitive terms to the mean values of the whole dataset. This way,
each cascade is classified  having either high social content or a low social
content, and having either a high cognitive content or a  low cognitive
content.

\begin{figure}[ht]
  \includegraphics[width=0.95\textwidth]{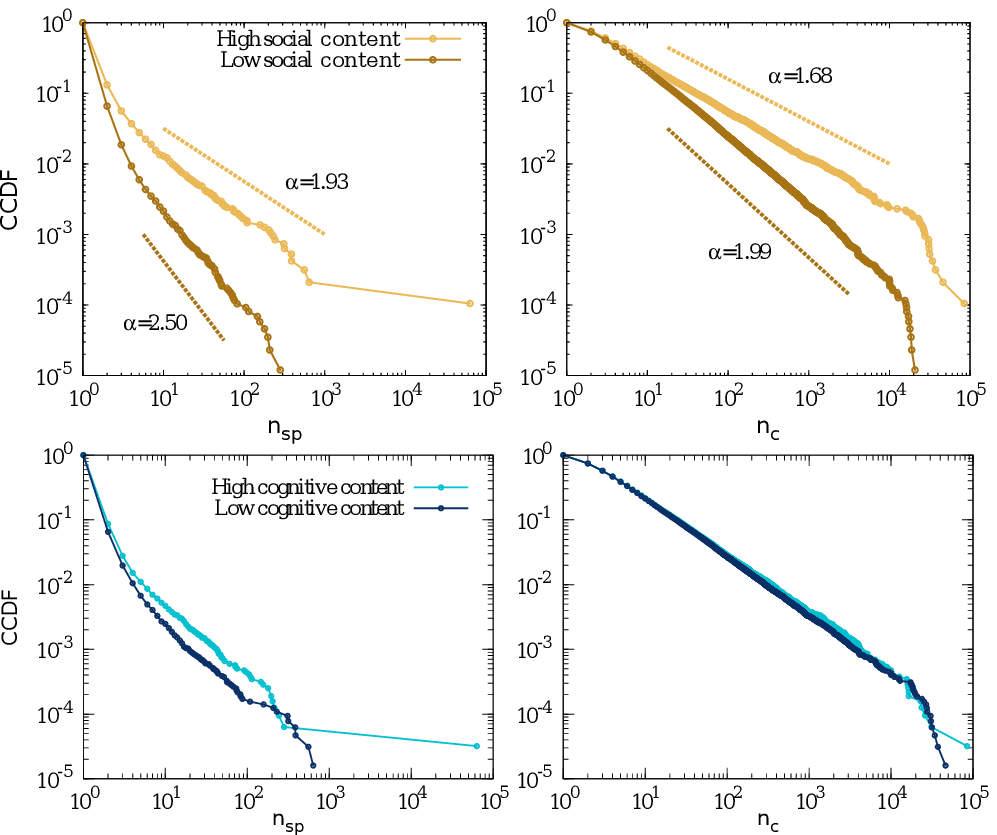}
  \caption{ CCDF of activity (left) and information (right) cascade
sizes for cascades of high and low  social content (top) and high and low
cognitive content (bottom). Dashed lines show the result of power-law fits. \label{fig:casc_LIWC_soc}}
      \end{figure}

The influence of social processes becomes evident when analyzing the
distributions  of cascade sizes depending on their social content, shown in
Fig~\ref{fig:casc_LIWC_soc}. The distributions of both information and
activity cascade sizes are different for high and low social content,  as
validated by a KS test (see SI Table IX). Power-law fits indicate that the
distribution of the size of activity cascades with high social content have an
exponent of $\alpha=1.87\pm 0.09$, while  the distribution for low social
content has an exponent of $2.33\pm 0.07$. This  difference highlights the role
of social processes in cascades during the formation of the  15M movement.
Cascades with social expression had an expected size that scaled with the size
of the movement, while those that did not include such language were
subcritical. For the case of  information cascades, the same result seems to
hold. In this case, however, information cascades with high social content
exhibit a power-law behavior with exponent $\alpha=1.66\pm 0.01$, which
indicates that the expected size scales with the system size. The outcome is
not so clear for low social content information cascades, for which
$\alpha=1.98\pm 0.02$ is compatible with $2$. However,  the latter are best
described by a log-normal distribution, as suggested by the log-likelihood
ratio $R$,  and the expected size of the audience does not scale  with the
system size (details on these fits can be found in SI Table X).

The above results indicate that the behavior of cascades (both of activity and
information) having high  social content is different from those where the
social content is lower.  On the contrary, words associated with cognitive
processes did not play such an important role in cascade sizes. The lower
panel of Fig~\ref{fig:casc_LIWC_soc}  shows the CCDF of cascade sizes
classified depending on their cognitive content. The cognitive content of the
tweets in an information cascade does not make it larger, as validated with a
KS  test (see SI). For the case of activity cascade sizes, a KS test rejects
the hypothesis that  they are the same, indicating that high cognitive content
have a slightly larger likelihood of  involving more spreaders, but not more
listeners. Power-law fits show that the exponents of  both types of cascades
are above $2$; while both exponents of information cascades are below $2$.

\subsection{The movement at the local level}

The above analysis shows how expression related to social processes and
emotions leads to spreading of activity and information through the social
network of 15M. The cascades present in the movement are not just large groups
of tweets; participants contribute repeatedly in these, and show heterogeneous
levels of engagement in the movement.

In this section, we test the principle of Durkheim's theory that social
integration in a movement leads to higher levels of participation, followed by
feelings of emotional synchrony with other participants in group actions.  The
main group actions of 15M were physical meetings in the center of towns,
demonstrations, and assemblies. But other kind of group activities took place
in the online medium. Tweet cascades created pockets of interaction within
\texttt{Twitter}, such that participants were aware of the large attention
that the movement was receiving online.  To quantify  the social activity of
each participant, we compute a vector of user features that quantifies the
integration in the movement, its level of activity, and its expressed emotions
and levels of social and cognitive content. We estimate participant
integration in the movement in terms of the follower/following network, i.e., 
a network in which a link from user $u$ to user $v$ is created when the latter 
follows the former. Thus, the direction of links goes from a user to its followers, 
indicating the direction in which information flows. 
We measure the $k$-core centrality of a user, $k_c(u)$
(explained in Materials and Methods), where the higher $k_{c}$, the better
integrated the user is. We also control for its amount of followers,
$k_{out}(u)$, and the amount of participants followed by $u$, $k_{in}(u)$. 
The level of engagement in the movement is approximated by the total amount of
tweets about 15M created by the participant, $n(u)$. We measure the expression
of emotions by means of the ratios of positive, $pos(u)$, and negative tweets,
$neg(u)$, and the ratios of words related to social processes, $soc(u)$, and
cognitive processes, $cog(u)$.

\begin{table}[ht]
\begin{center}
\begin{tabular}{l l l l l l l l l l l}
\hline
         & $n(u)$        & $k_c(u)$      & $k_{in}(u)$  & $k_{out}(u)$  & $pos(u)$    & $neg(u)$      & $soc(u)$       & $cog(u)$  &  R$^2$ \\ \hline          
$n(u)$   &               & $0.193^{***}$ & $0.015^{**}$ & $0.032^{***}$ & $0.010^{*}$ & $0.026^{***}$ & $-0.022^{***}$ & $-0.005$  &  $0.048$ \\  
$k_c(u)$ & $0.094^{***}$ &               & $0.676^{***}$& $0.090^{***}$ & $0.005$     & $0.012^{***}$ & $-0.012^{***}$ & $-0.003$  & $0.537$  \\ \hline
\multicolumn{5}{l}{\scriptsize{$^{***}p<0.001$, $^{**}p<0.01$, $^*p<0.05$}}
\end{tabular}
\caption{Linear regression results for individual activity level and integration in the movement. \label{table:individualLM2}}
\end{center}
\end{table}

\begin{figure}[ht]
  \includegraphics[angle=-90,width=0.95\textwidth]{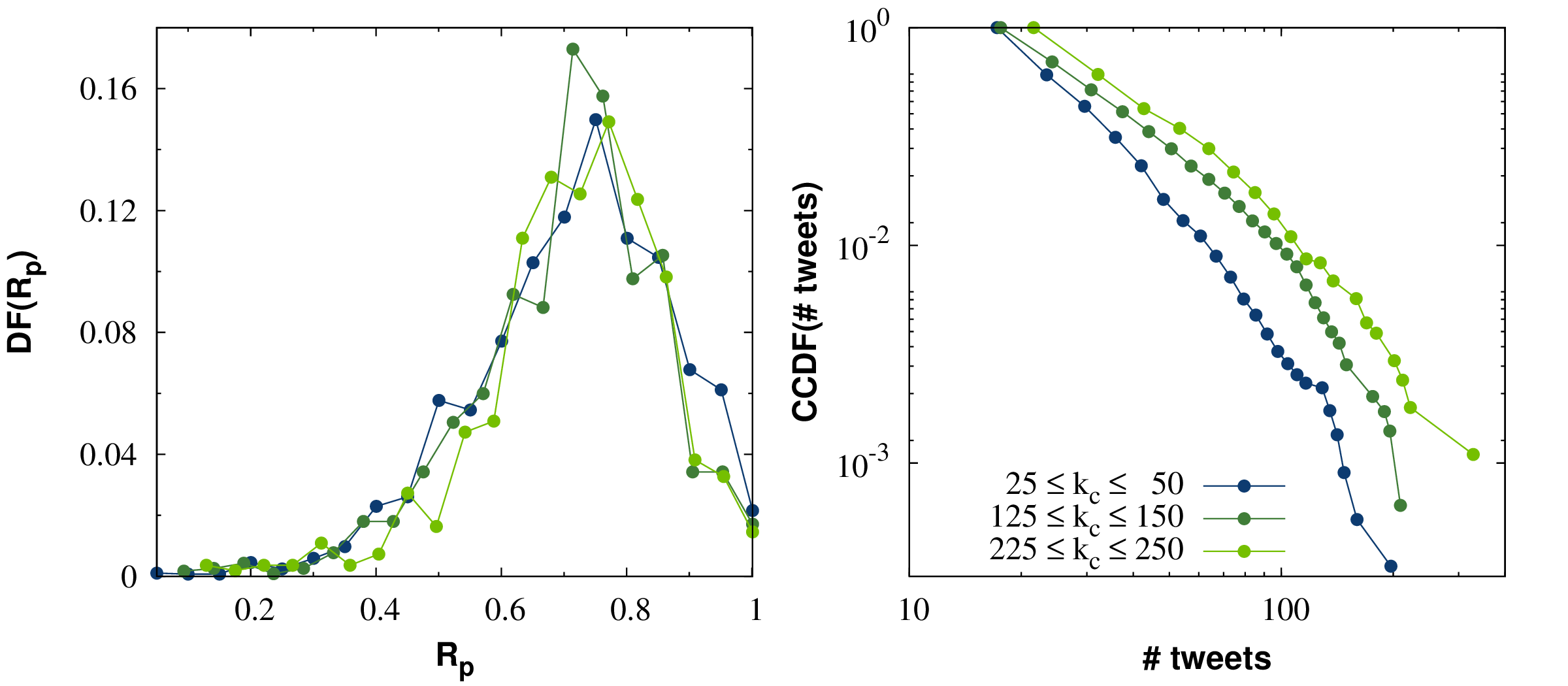}
  \caption{Left: probability density function of the ratio of
positive tweets of participants, for three ranges of $k_c$. Right: CCDF of the
engagement of participants, measured by their amount of tweets about 15M, for
three intervals of $k_c$. Participants with higher integration in the movement
are more engaged and active in the online medium. \label{fig:centrality_activity}}
 \end{figure}

We analyze the correlations between normalized versions of the variables using
a series of linear regressions. Table~\ref{table:individualLM2} shows the
results only for the regression of $n(u)$ and $k_c(u)$,  the rest is reported
in the SI Table XI. As proposed by Durkheim, the level of engagement in the
movement increases with social integration, estimated through the coreness of
the user. Other metrics, such as in- and out-degree, are also positively
related to the activity of a user, but with weight much smaller than the
weight of coreness.  The right panel of Fig~\ref{fig:centrality_activity}
shows the CCDF of $n(u)$ for different participants by their coreness.
Integration in the movement is correlated with activity, showing that
participants with higher integration in the movement are clearly more active.
It must be noted that this analysis does not test the causal nature of this
relation, but clearly rejects the null hypothesis of the independence between
activity and integration.

As visible in the left panel of Figure \ref{fig:centrality_activity}, positive
expression  has no significant effect on engagement. Participants with
different $k_c(u)$ do not have significantly different ratios of positive
tweets about 15M. Negative expression has a positive but small weight in both
activity and coreness, as reported in Table \ref{table:individualLM2}. We can
reject the hypotheses that negative expression is uncorrelated or negatively
correlated with integration in the movement, but the size effect of negative
expression in both integration and engagement is very low, as shown by the
weight in Table \ref{table:individualLM2}.

We do not find any significant relation between coreness and cognitive
expression. But we find a significant negative weight of social expression in
relation to both activity and coreness. This is consistent with our finding
that cascades with higher social content activate larger amounts of
participants. These cascades potentially start in the core and reach
participants with lower integration in the movement. Tweets about topics less related to
social processes do not reach the periphery of the movement. In conclusion,
less integrated users appear to be driven more by social processes.

An alternative condition for the emotions and the ratios of social and
cognitive terms of each participant is the assortativity with other
participants.  Members of the 15M movement  might be more emotional due to the
emotional expression of their immediate neighbors, in addition to their social
integration within the movement as a whole. To test this, we measure Pearson's
correlation coefficient of the ratios of positive, negative, neutral, social,
and cognitive tweet ratios of the participants, with the same ratio calculated
over the set of users that each participant follows inside the movement. We
replicate this analysis for two datasets, one based only on the tweets about
15M, and the other using an independent sample of 200 tweets per participant
(detailed in Materials and Methods). To test the significance of our results
against spurious correlations due to the network topology, we also computed
correlation coefficients in 1000 shuffled datasets in which the emotion,
cognitive, and social scores were permuted \cite{Fowler2008}.

\begin{table}[ht]
 \centering
 \begin{tabular}{|c | c c c c c|}
  \hline
    Dataset  &  $pos(u)$  &  $neg(u)$  &  $neu(u)$  &  $soc(u)$  &  $cog(u)$ \\
  \hline
 15M           &   0.063               & 0.068             & 0.065            & 0.035             & 0.128 \\ \hline
 15M shuffled  &   0.00002 (0.008)   & -0.0001 (0.007) & 0.0001 (0.007) & -0.0002 (0.008) & -0.0002 (0.007) \\ \hline
 individuals   &   0.261               & 0.364             & 0.315            & 0.336             & 0.358 \\ \hline
 ind. shuffled &   0.029 (0.01)     & 0.014 (0.009)    & 0.017 (0.009)    & 0.028 (0.009)    & 0.022 (0.009) \\ \hline
 \hline
 \end{tabular}
 \caption{Pearson coefficients of neighborhood correlations and means and two standard deviations of 1000 shuffled datasets. All coefficients have $p<10^{-10}$ in the empirical data. \label{tab:assort}}
\end{table}

Table \ref{tab:assort} reports the correlation coefficients for each dataset.
All correlations are significant and positive, indicating that the emotions
and semantic content expressed by a participant are correlated with  its first
neighbors, and thus emotions and psycholinguistic content are shared along
social links within the movement.  It should be noted that the correlation
coefficients for the 15M data are much weaker than for the 200 tweets from
each individual, indicating that the latter sample has more power to reveal
correlations in psycholinguistic analysis. While the 15M data is sparser and
noisier, the conclusions of the analysis of 15M data are consistent  with the
analysis of 200 tweets per individual, and robust with respect to the shuffled
datasets. Furthermore, these results are in line with previous research
\cite{Bollen2011} on emotional expression of subjective well-being,  and
extend the analysis with the presence of correlations for social and cognitive
terms beyond emotional expression in \texttt{Twitter}.

\section{Discussion}

The present work analyzes the evolution of the 15M movement through sentiment
and linguistic analysis of the participants' communication in the
\texttt{Twitter} social network. Using a dataset of tweets related to the 15M
movement, we track the activity of 84,698 \texttt{Twitter} users. Our
analysis includes 556,334 tweets during a period of 32 days, providing an
illustration of the structure of the movement in two ways: (i) at the dynamic
aspect of cascades in the discussion between connected users, and (ii) at the
individual level of social integration and participation of each user.

We combine psycholiguistics, sentiment analysis, and dynamic cascade analysis,
to understand the role of tweet content in the size and reach of collective
discussions in \texttt{Twitter}.  In line with previous works in social
psychology~\cite{Christophe1997}, we assess the role of emotions in social
interaction and collective action. We test the hypothesis that collective
emotions fuel social interaction by analyzing cascades according to their
emotional, cognitive, and social content.  We find that the sentiment
expressed in the first tweet of a cascade does not significantly impact the
size of the cascade. Instead, the collective emotions in the cascade are
responsible for its size in terms of spreaders and listeners. In particular,
cascades without positive content tend to be larger, and their size follows a
qualitatively different distribution. The cognitive content of the tweets of a
cascade play no role in their spread.  On the other hand, our analysis of
social content in the cascades reveals a clear pattern: cascades with large
ratios of social-related terms have distributions of listener and spreader
sizes that scale with system size,  in contrast with cascades with low ratios
of social-related terms, which follow distributions that have bounded means.

Our analysis at the individual level reveals that  users  are more integrated
in the movement, measured by their k-core centrality, if they exhibit higher
levels of engagement and express stronger negativity, in line with the overall
negative context of the movement (indignants). Our analysis also reveals that
highly integrated and influential users have  a lower tendency to express
social content in tweets. This indicates that social activation became salient
in the periphery of the movement rather than in its core. We emphasize that
our findings  are consistent with theories in sociology and social psychology
and confirm their statements by quantifying, for the first time, social and
psychological influence in collective action at large scale.

Our results have implications for research on social movements. The 
differentiation between social and cognitive processes is evident
when analyzing the size of cascades. Larger cascades have higher amounts of
social terms, invoking the participation of other users. This is also
consistent with our findings at the individual level: social  integration is
clearly  related to activity levels, showing the relevance of nonrational
factors in collective action. The members of a movement are not
deterministically defined by their demographic background and income. Instead,
the amount of social connections they have in the movement and their synchrony
with the emotions expressed by the movement as a whole are predictors of their
involvement.

Our findings show the added value of including additional psycholinguistic
classes into our analysis, i.e. the consideration of  social and cognitive
terms beyond sentiment analysis. Furthermore, our sentiment analysis
adaptation to Spanish demonstrate the relevance of sentiment analysis in
languages other than English, offering new opportunities to compare collective
phenomena in a wide variety of societies and political systems.

Beyond social movement analysis, our work has implications for studying other
online phenomena, such as memes or viral marketing campaigns. Our findings on
cascade sizes for different psycholinguistic classes suggest that  words
related to social processes  lead to  larger collective responses in
\texttt{Twitter}, pushing the virality of content above a critical threshold
that produces qualitatively different cascading behavior.

\section{Materials and Methods}

\textbf{15M tweets and network.}  Our dataset comprises activity from
\texttt{Twitter} related to the 15M movement in Spain, which brewed for some
time in several online social media, and mainly rised with the launch of the digital platform 
\textit{Democracia Real Ya} (\textit{Real Democracy Now}). Twitter and Facebook were utilized to 
organize a series of protests that took off on the 15th of May, 2011, when demonstrators camped in several 
cities \cite{borge2011,borge2012}.  From that moment on, camps, demonstrations and protests 
spread throughout the country, and the 15M became a grassroots movement for 
additional citizen platforms and organizations. 
As many of the adherents are online social media users, the growth and stabilization of the movement was 
closely reflected in time-stamped data of twitter messages. Some of these tweets were extracted from the
\texttt{Twitter} API according to a set of pre-selected keywords (see SI Table
I), and the collection comprises messages exchanged from the 25th of April 
at 00:03:26 to the 26th of May at 23:59:55, 2011.  The sample of tweets was filtered by the Spanish startup company
\textit{Cierzo Development LTd.},  which exploits its own private SMMART
(Social Media Marketing Analysis and Reporting Tool) platform,  and therefore
no further details are available. According to previous reports, the SMMART
platform collects 1/3 of the total \texttt{Twitter} traffic. From  the sample
of tweets we obtained, the follower/following network is  extracted: for the
active users, i.e. those who posted at least one tweet in the sample
collected, the set of  followers is retrieved, and the resulting network is
filtered to include only the active followers. The resulting network is
composed of nodes that represent users, and edges with directionality
corresponding to the information flow in \texttt{Twitter}. This way, if a user
$u$ is a follower of user $v$, there will be a directed link from $v$ to $u$
in the network.

\textbf{Sentiment analysis.} To detect the sentiment expressed in each tweet,
we apply the Spanish adaptation of \texttt{SentiStrength} \cite{Garcia2013c},
a state-of-the-art sentiment analysis tool for short, informal messages from
social media \cite{Thelwall2013}. \texttt{SentiStrength} is used in a wide
variety of applications, from the sentiment analysis of stock markets
\cite{Zheludev2014}, to reactions to political campaigns \cite{Garcia2012},
and interaction in different social networks \cite{Thelwall2013}. We tailored
\texttt{SentiStrength} to the Spanish language based on a sentiment corpus of
more than $60000$ tweets and evaluated it on an independent corpus of more
than $7000$ human-annotated tweets \cite{Diaz2013}. More details about our
application of \texttt{SentiStrength} and the results of this evaluation can
be found in the Supplementary Information. After sentiment detection, for each
tweet $m$, we have an emotion value $e_m$ associated with the tweet.
$e_{m}=1$ if the tweet is positive with respect to its emotional charge,
$e_{m}=0$ if the tweet is neutral, and $e_{m}=-1$ if the tweet is negative.
We abbreviate these as \emph{positive}, \emph{neutral} and \emph{negative}
tweets, always referring to their emotional charge.

\begin{comment}
\textbf{News data.} 
Media coverage was assessed by querying the database Nexis and Google News for
headlines in Spanish for three main keywords linked to the protests ("15-M" or
"indignados" or "Democracia Real").
\end{comment}

\textbf{Linguistic content analysis.} We analyze the content of tweets based
on frequencies of terms from the Linguistic Inquiry and Word Count lexicon
(LIWC) \cite{Chung2011}.  This lexicon is a standard technique for
psycholinguistics, including terms associated to affect, cognition, and social
processes. LIWC has been used to predict suicides \cite{Stirman2001}, and to
analyze collective mood fluctuations \cite{Golder2011}. For each tweet, we
apply a simple dictionary detection technique based on the lemmas of the
lexicon, stemming the tweets and detecting the use of terms in the LIWC
classes of social and cognitive processes. This way, for each tweet $m$ we
have two counts of social, $soc_m$, and cognitive, $cog_m$, terms as well as
the amount of words,  $w_m$, in the tweet.

\textbf{Cascade detection.} Cascades in online networks may be defined in
several ways with respect to the variety of  online platforms, discussion
topics, or interaction means between users. Here we adopt the definition first
described in  \cite{borge2012} which is based on \textit{time-constrained}
cascades. Time is discretized according  to a window width, and tweets posted
at consecutive  time windows are considered to be part of the same piece of
information if users emitting them show a follower/following relationship. 
Specifically, user  $u_j$ posting the tweet at time $t_1$ must follow user
$u_i$ sending the tweet at time $t_0$. Previous works showed the robustness of cascade statistics for different time 
windows \cite{borge2011, borge2012}. Here we choose a $24$ hours-window to 
minimize the eventual correlations due to the effect of circadian activity 
in human online behavior. The  content being sent is not required
to be the same. This is motivated by three main facts: first, our sample,
i.e., the set of tweets,  has been previously filtered by topic, allowing us
to safely assume that the information circulating is limited to a restricted
topic, the 15M movement. Second, the 15M movement is a deliberative process,
characterized by discussions and debates about the  political and social
situation of the country, the organization of protests and demonstrations, and
conversations about the strategies to  follow. Finally, \texttt{Twitter} is
considered to be both a micro-blogging service and a message interchange
service, as  suggested by the high values of link reciprocity $\rho \sim 0.49$
and the \texttt{mention} functionality.  Time-constrained cascades allow to
take into account these frequent situations  in which people discuss about
particular topics using their own words to express their ideas, rather than
forwarding a restricted piece of  information.

A cascade is then an ordered set of consecutive activities of a set of users
having follower relations. This way we know who started the cascade and when,
and the seed tweet triggering the  cascade. We can additionally distinguish
between activity cascades and information cascades. The  first ones involve
only the set of active users, i.e. those responding to the message, whereas
information  cascades also comprise listeners, i.e. users receiving the
message but not participating in the discussion. We want to first investigate
if the initial tweet determines the size of  the cascade. For instance, one
could argue that positive messages can trigger larger cascades, or  vice
versa, negative messages trigger a debate that can last over several time
windows.

\textbf{k-core centrality.}
The $k$-core value is an individual measure of importance based on the core
structure of the network.  A $k$-core is defined as the largest subnetwork
comprising nodes of degree at least $k$. Note that to compute this measure 
we consider the undirected network, i.e., we consider every link as if 
it was undirected, and therefore a node of degree $k$ is a node whose total 
degree is $k=k_{in}+k_{out}$. The $k$-core decomposition method
assigns an integer number to every node in the network, obtained by a
recursive pruning of their links. The procedure  starts with isolated nodes,
which are assigned a $k$-core value $k_c=0$. Then, nodes with degree $k=1$ are
removed along with their links,  and assigned $k_c=1$. If any of the remaining
nodes is left with $k=1$ connections it is also removed and contained in the
$k_c=1$ core. The process continues with $k_c=2,3, \dots$ until every node has
been assigned to a $k_c$ shell. This metric  goes beyond degree, as it takes
into account the centrality of the neighbors to define the centrality of a
node.

\textbf{200 tweet timeline data.}
An extended dataset of tweets from the participants was also extracted. It
comprises the last 200 tweets (if available) posted by the set of active
users. It consists of a sample of 15,411,025 tweets (see SI Table IV)
retrieved the 20th of October, 2013.

\begin{backmatter}

\section*{Competing interests}
  The authors declare that they have no competing interests.

\section*{Author's contributions}
RA and YM gathered data; RA and DG analyzed the data;
RA, DG, YM, FS designed research and wrote the article.

\section*{Acknowledgements}
DG and FS acknowledge financial support by the Swiss National Science Foundation (CR21I1\_146499).
RA, FS, and YM acknowledge financial support by EU-FET project MULTIPLEX 317532.

\bibliographystyle{bmc-mathphys} % Style BST file
\bibliography{AGMS}      % Bibliography file (usually '*.bib' )

\section*{Additional Files}
  \subsection*{Supplementary Information}

\end{backmatter}
\end{document}